\documentclass[aps,prl,twocolumn,showpacs,superscriptaddress]{revtex4}   % for review and submission with two columns
\usepackage{graphicx}  % needed for figures
\usepackage{dcolumn}   % needed for some tables
\usepackage{bm}        % for math
\usepackage{amssymb}   % for math
\usepackage{verbatim}  % for comment
\usepackage{epstopdf}
\usepackage{color}
\usepackage{soul}
\usepackage{amsmath}
\usepackage{subfigure}
\usepackage{color,colordvi,graphicx,pstcol}

\newcommand{\beq}{\begin{equation}}
\newcommand{\eeq}{\end{equation}}
\newcommand{\bea}{\begin{eqnarray}}
\newcommand{\eea}{\end{eqnarray}}

\newcommand{\commentout}[1]{{}}

\begin{document}

\title{Finding Optimal Training Parameters for Quantum Generative Adversarial Networks}
% Above title was Susanne's suggestion.
\author{C. Strynar}
\affiliation{Department of Physics, University of Massachusetts,
Dartmouth, MA 02747}
\author{R. M. Rajapakse}
\affiliation{Department of Physics, University of Massachusetts,
Dartmouth, MA 02747}
\affiliation{Department of Physics, University of Connecticut,
Storrs, CT 06269}

\date{\today}

\begin{abstract}
Some of the most impressive achievements of contemporary Machine Learning systems comes from the GAN (Generative Adversarial Network) structure. DALLE-2 and GPT- 3, two of the most impressive and recognizable feats of ML in recent years, were both trained using adversarial techniques. The world of Quantum Computing is already well aware of the value of such techniques on near-term Quantum Hardware: QGANs provide a highly efficient method for loading classical data into a quantum state. We investigate the performance of these techniques in an attempt to determine some of the optimal training parameters in a Qiskit-style Parameterized Circuit QGAN framework.
\end{abstract}
\maketitle
\section{Introduction}
Generative Adversarial Networks are one popular paradigm of contemporary machine learning (ML) for generating data. Some of the most recognizable ML systems of the past few years have been trained using adversarial techniques \cite{arxivQuantumGenerative, LiZhang, BiamonteWittek, Ngo}. ML systems have been used in agriculture\cite{Holmes, Jeong, Besombes, Pantazi}, finance\cite{LussangeLazarevich}, healthcare\cite{HazraKumar}and education\cite{DjambicKracjar, HuangDu, Holmes}. In this paper we will first build up a basic understanding of the technical concepts required to understand Quantum Generative Adversarial Networks (QGAN), which include an overview of quantum algorithms and classical Generative Adversarial Network (GAN) architecture. We will then discuss the possible future uses of QGANs on real quantum hardware to motivate the proceeding investigation on their complexity and performance. We find that QGANs have a computational complexity that does indeed suggest their viability for real world data-loading on quantum computers\cite{arxivQuantumGenerative,arxivQuantumGenerative1, arxivVariationalQuantum, qiskitQuantumNeural,arxivUnification, Chuang, Preskill}. We conclude with a discussion of the possible future needs of quantum algorithm users, and which research directions should lead to the most valuable information for these future users. 

To describe it using formal math concepts, a GAN tries to solve the following problem setup: given a "data format" $U$ and a dataset $X\subset U$, find an algorithm $D(x): U \rightarrow \{0, 1\} \approx x \in X$, and an algorithm that parameterizes this subset $G(z): Z \rightarrow U$ such that the image $Im(G) \approx X$ \cite{arxivQuantumGenerative, arxivQuantumAmplitude, HarrowMontanaro}.

Concretely, we might employ a GAN to "learn" the dataset of all images of cats. In this case, $U$ is the space of all possible images, and $X \subset U$ is the set of all images of cats. In other words, given an image $x \in U$, $x \in X$ if and only if $x$ is an image of a cat. We would like to determine if a given image is an image of a cat, so we would like to find a function $D$ such that $D(x)$ tells us whether or not $x \in X$. We would also like to be able to generate instances of $X$, i.e. given a random input $z$ we would like to find an onto function $G$ such that $G(z)$ gives an $x \in X$, and every $x$ has a $z$ such that $G(z) = x$\cite{HuangDu}. 

In reality, given $U$, $D$ is unique, and given a parameterizing space $Z$, $G$ is unique down to an isomorphism of $Z$. However we are not concerned about finding these functions exactly. Notice the $\approx$ in the problem setup. We will have succeeded so long as we find a $D$ and $G$ that are "good enough", as is often the case in machine learning. This reduction turns out to drastically reduce the computational cost of the problem \cite{arxivVariationalQuantum, HuangDu}.

With the abstract problem structure in mind, we attempt to find a solution using a GAN \cite{VintAnderson}:

\includegraphics[scale=0.3]{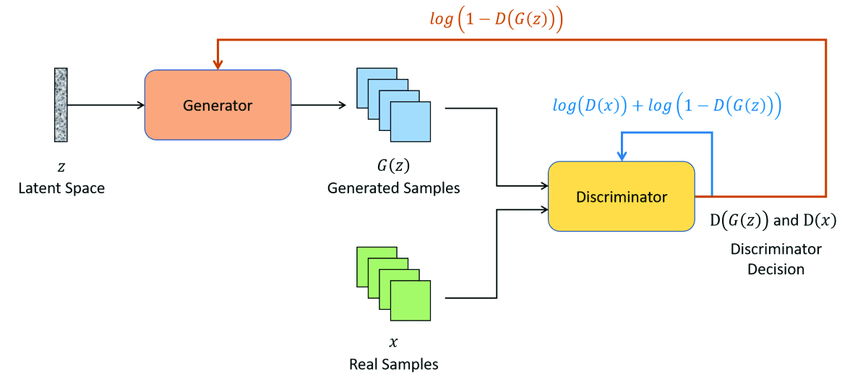}
    \emph{Fig 1: Standard GAN Architecture}

Figure 1 illustrates the standard structure of a GAN. The important features are the "Generator" and "Discriminator" blocks which each initially implement a random function. Eventually, we would want these functions to be our functions $G$ and $D$. The system first generates a handful of samples, and select a handful of real samples. These are all passed through the discriminator, and the discriminator is then rated on on how well it classifies the real samples as real, and the generated samples as fake. The measure how well the two algorithms performed are known as the "loss functions", which are defined by the logarithmic expressions in Figure 1 (blue for the discriminator, red for the generator). Finally, the generator and discriminator are updated using their performance according to the loss function to slightly change the algorithms for the better \cite{arxivQuantumGenerative,apsQuantumstatePreparation}.

Before being concerned about how this updating of algorithms happens, let us assume it generates a result and study the behavior of the problem. It can be shown that in the long run, the generator will perfectly generate the desired dataset $X$ and the discriminator will have no better choice but to randomly guess "real" or "fake" on all inputs. This state of affairs is often called the Nash equilibrium of the system \cite{arxivQuantumGenerative}. A multi-agent system is in a Nash equilibrium when no single agent can improve their reward by changing only their own actions assuming the constancy of everyone else's actions. In the present case, a GAN is a "multi-agent interaction" between the generator "agent" and the discriminator "agent". The aforementioned situation is a Nash equilibrium because the generator can do no better than perfectly generating the dataset, and the discriminator can do no better than randomly guessing. In any other situation, the generator can always do better by coming closer to generating the dataset perfectly, and the discriminator can always do better by trying to exploit the specific biases of the current generating algorithm. Since the only stable state is the one where the generator has perfectly replicated the dataset, in the limit of infinite time this structure must reach this equilibrium, and success is assured. 

If the algorithms are neural networks, back-propagation could be done using the loss function, and updating the network's weights using gradient descent, as is a standard practice in machine learning.

The general idea here, and indeed in most of machine learning, is that algorithm-space has to be parametrized. Writing algorithms directly is not attempted, but we attempt to write an algorithm that will \emph{find} an algorithm. Parameterizing algorithm-space is necessary for this. If a correspondence between $\mathbb{R}^n$ and possible algorithms is set up, we can begin at some $\vec{v} \in \mathbb{R}^n$, check the accuracy with which the algorithm solves our desired problem, and then slightly change $\vec{v}$ in a direction that improves the algorithm. Repeat this process for a while and theoretically one would obtain at a vector $\vec{v}$ that corresponds to a well-performing algorithm. This is exactly what standard neural networks do: the weights (along with the number of layers, width, and activation function) define the algorithm, and over time back-propagation is used on the loss function to update these weights. 

These specific points are worth mentioning, as they are important in the following discussion of quantum GANs.

\section{Review of Quantum Algorithms and Parameterized Quantum Circuits}

For reasons that will be discussed later, it turns out to be a useful thing to use a GAN to "learn" quantum datasets. The most efficient way to do this is to use quantum algorithms for both our generator and discriminator blocks. Given an oracle that contains some quantum distribution, we'd like our generator to learn to replicate this distribution. Much like we had to come up with some way to parameterize classical algorithms in order to do machine learning, such as using a neural network, we will also need to find some way to parameterize \emph{quantum} algorithms. 

One common way of doing this is by using parameterized quantum circuits, or PQCs. The idea is that there are quantum gates that already come parameterized by a real number, such as a rotation gate. 

\begin{center}
    \includegraphics[scale=0.4]{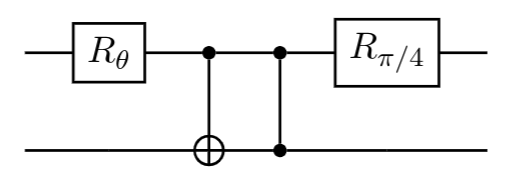}
    
    \emph{Fig 2: Basic parameterized circuit with a rotation gate, parameterized by $\theta$}
\end{center}

This is convenient; all that needs to be done is to put rotation gates together, leave the rotation angles unspecified, and thereby successfully parameterize a part of quantum algorithm-space. By specifying a list of rotation angles, each combination would correspond to a particular quantum algorithm, which is exactly what is needed. This is a parametrized quantum circuit (PQC). Given a loss function, similar to what is done with classical machine learning systems that parameterize classical algorithms with real parameters called weights, gradient descent is used to iteratively update our PQCs in a promising direction, until a quantum algorithm that performs well according to the loss function is obtained. 

It is useful to differentiate between the \emph{structure} of a PQC, versus any particular circuit that has the rotation angles filled out. A general layout of unspecified rotation gates is known as a \emph{circuit framework}, whereas specifying the rotation angles of a framework gives a specific \emph{circuit}.

Addressing the question of picking a circuit framework, the circuit diagrams for the layouts we have decided to use for our two-qubit generator and discriminator$^3$ are given below:

\begin{center}
    \includegraphics[scale=0.3]{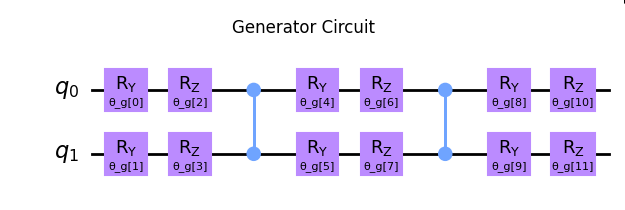}

    \includegraphics[scale=0.3]{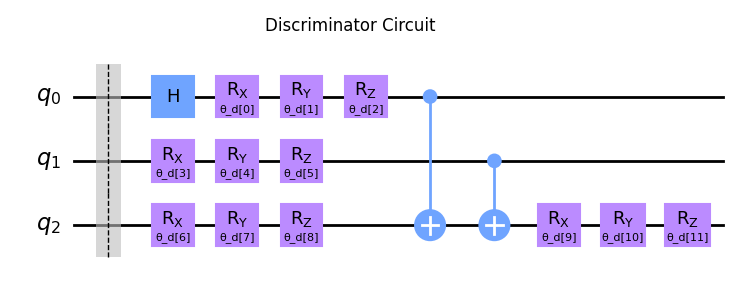}

    \emph{Fig 3: PQCs used as the generator and discriminator structures. }
\end{center}

These are not quantum circuits yet, because the rotation angles are not specified. In practice, each gate will have a rotation angle. As we train the system, the rotation angles slowly update in a "good" direction, but the layouts themselves \emph{do not change}. Every 2-qubit generator and discriminator circuit used takes these particular forms, even if the rotation angles are not the same.

What makes for a good circuit framework? One measure worth talking about is \emph{expressibility}. Expressibility is a way of talking about a given circuit framework's algorithm-space coverage. In trying to find a satisfactory quantum algorithm, it is necessary to make sure that a reasonable approximation will appear in the search space. 

Tailoring the circuit framework to the particular problem being solved may be desirable, but is typically difficult, and will sacrifice generality. If the circuit framework has high algorithm-space coverage there is a good chance that a satisfactory algorithm may be found. This is the point of using highly expressible circuit frameworks: there is a sort of almost-guarantee that the training process will succeed eventually.

Now to determine if the setups are "expressible enough", the size of quantum algorithm space is discussed.

The 1-qubit algorithm space is simple : all  that is needed is a U-gate, which takes three real parameters and gives a 1-qubit algorithm. 

\begin{center}
    \includegraphics[scale=0.7]{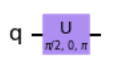}

    \emph{Fig 4: U-gate drawn with the Qiskit python library}
\end{center}

This circuit above is the most general algorithm mapping 1-qubit to 1-qubit. Indeed, this is the whole point of U-gates, the "U" standing for "universal". In the figure the U-gate has specific parameters $\frac{\pi}{2}, 0, $ and $\pi$, which is a Hadamard gate. These three numbers can take on any value in $[0, 2\pi)$, and each combination gives a different algorithm. Thus we say that 1-qubit algorithm space is 3-dimensional: it takes three independent real parameters to specify each 1-qubit algorithm uniquely. 

2-qubit algorithm space is much more rich in structure. It should not be surprising that the space is more than 6-dimensional. Here is a fully-general 2-qubit circuit:

\begin{center}
    \includegraphics[scale=0.4]{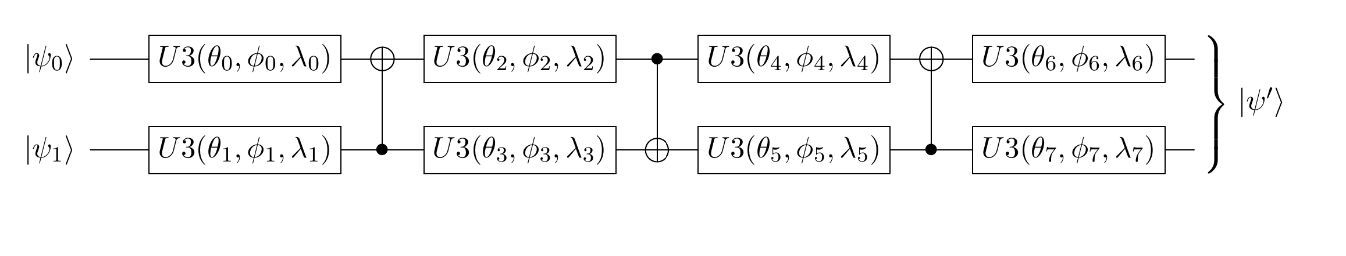}

    \emph{Fig 5: A fully-general 2-qubit algorithm, taking 8 independent U-gates, and 3 CNOT gates}
\end{center}

2-qubit algorithm-space is \emph{24 dimensional}! It takes 24 independent real parameters in $[0, \frac{\pi}{2})$ to uniquely specify a 2-qubit algorithm. I'm unsure of where exactly this number comes from, or how the size of quantum algorithm space will grow with the number of qubits. It's size does have to be bounded by $4^{n+1}$ dimensions, because at most $n$ qubits are represented with $2^n$ complex numbers, i.e. $2^{n+1}$ real numbers. A linear mapping from $\mathbb{R}^{2^{n+1}} \rightarrow \mathbb{R}^{2^{n+1}}$ is a $2^{n+1} \times 2^{n+1}$ matrix, which has $(2^{n+1})^2 = 4^{n+1}$ real entries. 

This limit says that a 1-qubit algorithms can be at most $4^{(1)+1} = 4^2 = 16$ dimensional, and that 2-qubit algorithms can be at most $4^{(2)+1} = 4^3 = 64$ dimensional. These are quite off from the actual dimensions of 3 and 24, because we have not yet factored in conditions such as unitarity and global phase equivalence. 

In order to have a decent hope of finding a good algorithm in n-qubit algorithm space, we need to make sure that the number of parameters of our circuit frameworks are a decent size portion of the size of n-qubit algorithm space. Both of the frameworks we have chosen for 2-qubits have 12 independent parameters. While this may seem quite far off from the theoretical ideal of 24, it turns out to be good enough for our purposes, because of a simplification that will be made in the next section, and we are able to achieve an accuracy of $10^{-4}$ KL-Divergence, measured in nats (the logarithms use base $e$). This is noticeably worse than the accuracy achieved for 1 qubit, where we use the full range of 1-qubit algorithm space to achieve an accuracy in some cases down to machine round-off error. 

Had we desired to achieve accuracy to machine round-off error for 2-qubits or higher, we would have used larger circuit frameworks and spent more compute on the training phases. For many practical purposes however, our accuracy is fine. There is a choice to be made though. As is often the case, there is a trade-off between computational cost and accuracy. As will be discussed later, investigating different frameworks and the best way to make this decision of cost vs accuracy based off of your specific problem is a direction of future research that holds promise to significantly help future quantum users.

Now that we have established some of the basic theoretical building blocks we need to understand quantum GANs, we will now justify their use. 

As discussed previously, quantum GANs allow us to learn an arbitrary quantum algorithm given a loss function. Instead of discussing the algorithm we learn in full generality, we can decide to look at how the algorithm transforms \emph{only} the $|\vec{0}\rangle$ state. In other words, we only care about how the algorithm transforms one very specific state, instead of worrying about how it transforms any arbitrary state. Instead of learning a quantum algorithm, it's more like we're learning a particular quantum state.

This is a big part of why we are able to do so well despite only covering about half of quantum algorithm-space (remember we use 12 out of a theoretical 24 parameters for 2-qubits). We only care about how the zero state is transformed, which leads to some redundancy in algorithm-space, i.e. a number of different algorithms will be identical to us for our purposes. So the chances that we are able to find \emph{any one} of those "best" algorithms is a lot higher than the chance that we'd be able to find a particular best algorithm. This also means that using the full size of algorithm-space is unnecessary, even if you'd like to learn a perfect algorithm. 

So making this simplification allows us to learn a quantum state, or more accurately we learn a loading algorithm that transforms the zero state into our desired state, but the question remains. Why is this something that we might want to do?

This is a serious problem if we would like the use of quantum algorithms to be feasible.This is the purpose that QGANs hope to fill: to provide a faster way (hopefully non-exponential) to load data. 

The sort of QGAN used depends on the type of data being loaded. For loading quantum data, both the generator and the discriminator will be quantum circuits. If the data being loaded is classical, then a hybrid system will be used where the generator is quantum but the discriminator is classical. Either way we are generating a quantum state (otherwise there would be no point in using a \emph{quantum} GAN), but sometimes we would like to learn a quantum representation of a classical dataset, instead of learning a quantum distribution. 

The difference is where the measurement takes place. With a quantum discriminator, the output of the generator (or the real distribution provided by an oracle) is passed into the discriminator which processes that signal using quantum gates, which is then measured and used to calculate the loss function and update the parameters of each system. With a classical discriminator, the output of the generator (or oracle) is measured \emph{immediately}, and the output is passed into the discriminator and gets processed with classical operations (perhaps via a neural network), and then the output is once again used to calculate loss and update the parameters via gradient descent. We will only concern ourselves with a fully quantum QGAN, so our discriminator is a quantum circuit. Hybrid QGANs are a direction for future research. 

It should also be noted that \emph{all} QGANs still rely on classical computation. The parameters are still updated with gradient descent, which is calculated classically, even though the parameters correspond to quantum circuits. There are some approaches where machine learning is done solely using quantum computation, but that is not what we are doing here.\section{Setup}

First, we will learn a 2-qubit Bell-state. Before we do any training, the parameters of the generator and discriminator are randomly initialized, which is a common practice in machine learning. Here is the randomly initialized state of the generator, along with the Bell-state we would like to learn:

\begin{center}
    \includegraphics[scale=0.3]{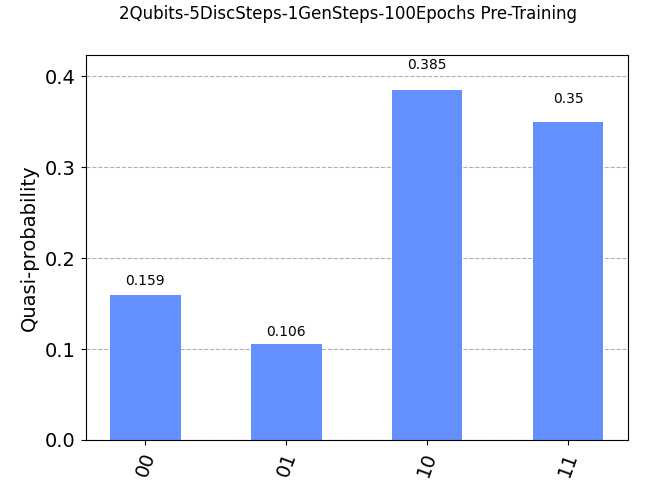} \includegraphics[scale=0.4]{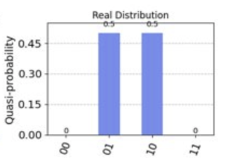}

    \emph{Fig 7: randomly-initialized state of the generator (let), Bell-state we are trying to learn (right)}
\end{center}

Now before we can start training, we have to build our circuits. The 2-qubit circuit layouts for our generator and discriminator are shown in figure 3, but these are not used alone: we have to put them together to get one larger circuit.

There are two different circuits that will be used:

\begin{center}
    \includegraphics[scale = 0.3]{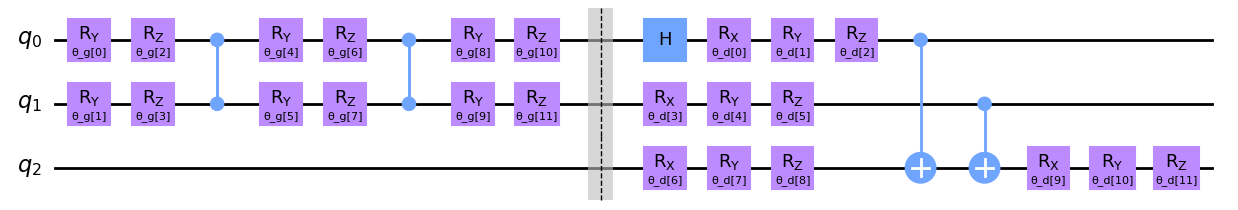}

    \includegraphics[scale = 0.3]{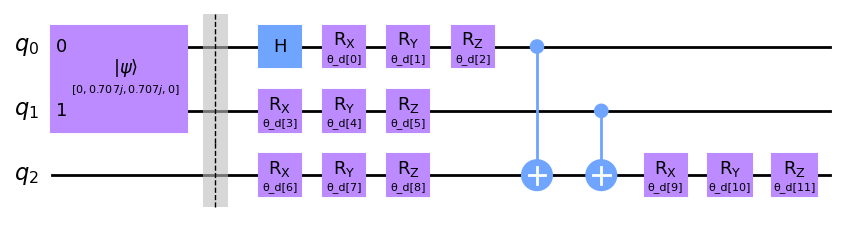}

    \emph{Fig 8: concatenated quantum circuits for a 2-qubitBell-state training run}
\end{center}

The top circuit is our generator circuit and our discriminator circuit put together into one completed circuit. During training, a measurement will be performed on the bottom-most qubit, $q_2$, the result of which will be used to update parameters. On the bottom circuit, we have our discriminator circuit put together with a quantum oracle that holds the Bell-state we are trying to learn. Once again, training will involve taking a measurement of the bottom-most qubit, and updating weights accordingly.

During training both of these circuits will be used. The top circuit is the one where the discriminator learns to tell apart the generator from the real distribution, and  the bottom circuit is where the discriminator learns to recognize the real distribution. Use of the second without the first would train our discriminator, but not our generator. Use of the first without the second would train our generator and discriminator against each other, but there would be nothing to nudge them into learning our desired Bell-state, they would just try to outdo each other with no aim. Run them together though, and learning our Bell-state is the best option for them in the interest of their programmed loss functions. 

So that one circuit does not have more of an influence than the other, both are used together when updating the discriminator. This does not need to be the case. Perhaps it would be more efficient to do something else. Investigating this is one possible direction for future research. 

\section{Training Process}

Now to talk about the training process. There is a bit of variety for things we can do. It is helpful to think of "generator updates" and "discriminator updates" as the "atomic units" of the training process, and so we discuss these first.

\subsection{Atomic Updates}

One generator update only involves use of the top circuit in Fig. 7. We have current parameters for both circuits, and they are loaded in. A classical simulation is used for computing the final measurement of the circuit. The use of a classical simulation is where much of the computation cost of these experiments comes from - and is the reason we are only able to run trials with a small number of qubits. This will be discussed more later.

Once the measurement is made, this tells us "how well" the current weights of the discriminator did against the current weights of the generator. To update the generator though, we need to know how well the generator did, and as is clear in the general loss function formulas for a GAN in figure 1, we take how well the discriminator did and "invert" it to determine how well the generator did. 

Now that we have computed the loss of a given set of generator parameters, we can use a built-in qiskit function to calculate the gradient, and a standard optimizer to update the parameters in a worthwhile direction. There are two things that define the updating process (beyond the current set of weights), and that is the chosen optimizer function, and the learning rate. The learning rate can be thought of as "how fast the parameters get updated". It may sound like you'd always want a high learning rate, but if you update your parameters too fast you risk jumping over the optimal parameters you're looking for. We use a learning rate of $0.02$. The specific optimizer that is chosen determines the exact manner in which the parameters get updated given the gradient. A very common optimizer to use in machine learning is the Adam optimizer, as it tends to achieve efficient performance across a wide range of tasks. Testing out different combinations of optimizer choice and learning rate was not something we concerned ourselves with, and its a worthwhile direction for future investigation in the hopes of further improving computational efficiency.

All that is what one "generator update" is. A simulator determines the outcome of the measurement of our first circuit, using the current parameters of both the generator and the discriminator. The outcome determines the loss, a qiskit function determines the gradient, and finally the Adam optimizer determines the new set of generator parameters using this information. 

One discriminator update is similar, but actually does this process twice, once with each circuit. It simulates both circuits \emph{at the same time} (using the current generator/discriminator parameters), calculates the parameter updates for each circuit individually, and then applies the changes all at once. 

\subsection{The Training Process}

The training process will involve running some series of these updates, after which we hope we will have found some parameters that perform decently. 

In order to achieve a satisfactory outcome, it is important to train both systems together. If we were to train them completely in sequence, it can be seen that things will go wrong. Consider that we train the discriminator all at once, before we then train the generator. Pretty quickly the discriminator will learn to simply answer "real" no matter what it gets! It would not learn when to say "fake" at all. Considering the other way, if we trained the generator first and all at once, it would never learn our distribution because it would never come in "causal contact" with it! The generator only sees that discriminator, which starts as having randomly-initialized parameters. While the generator would probably learn how to get the discriminator to say "real", it wouldn't matter because this discriminator is completely uncorrelated with our distribution. 

So training our systems together is important. One way this can be done is through a series of training "epochs". In each epoch, we use a predetermined number of generator and discriminator updates. For instance, in each epoch we might update our discriminator 5 times and our generator once. We then use this same sequence throughout the duration of our training. We call the ratio of discriminator to generator updates in each epoch the "training ratio", because in our tests we keep the training ratio constant across the entirety of the training session. 

This is not the only way to do it. There's no real reason that the ratio of generator and discriminator updates needs to stay constant over the course of all the training epochs, other than that it makes things simpler. Perhaps it could be more efficient to begin with a higher training ratio at the beginning (more discriminator updates), and then switch to a lower training ratio towards the end (more generator updates). We did not worry about these possibilities, and kept things simple with a constant training ratio. This is yet another interesting possible direction for future research. 

\subsection{Verifying Solutions}

One last thing before we get back to our example of learning a Bell-state. We need some way of measuring "how well" we've done, outside of just using the loss functions. As was illustrated in the examples given where the circuits are trained entirely in series, very low training loss was achieved, but our system did not get anywhere remotely close to achieving what we wanted it to achieve. Even in normal settings, it's possible that our system achieves a low loss undeservedly. In order to make sure we aren't deluding ourselves, it's important that we check how well our system is actually doing, by checking how "close" the output distribution is to the desired distribution. Kullback-Liebler divergence is a standard way of measuring the "difference" between two probability distributions, and so is a natural measure to use for our purposes. 

KL divergence is well-defined for two probability distributions (order matters - KL divergence is generally not commutative) down to a choice of base for the logarithms. Use of base-2 logarithms means that KL divergence takes on units of "bits". We use the natural logarithm, giving our KL divergence units of "nats". Here is the formula used to calculate it in the discrete case (our purposes are discrete):

\begin{center}
    \includegraphics[scale=0.7]{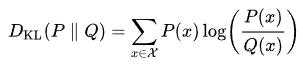}

    \emph{Fig 9: Calculation of KL-Divergence for a discrete distribution}
\end{center}

One way to think of KL divergence is as the "relative entropy" or "information gain" of P over Q. For our purposes, we'd like to check if the distribution outputted by our generator matches our desired distribution, meaning we'd like the KL divergence between the two distributions to be as close to zero as possible. 

You may wonder why we bother using a loss function that does not perfectly match the thing we want. In many cases in machine learning, this is done because there is no precise way to quantify the exact thing we want, and so a proxy is used. But I just finished explaining a precise way to measure the thing we actually want, so why not use it? In fact, why even bother using a QGAN at all? Why not just train the generator alone, using the KL divergence as a loss function? 

The answer is that in real life we \emph{do not have direct access to the real distribution}, which is a crucial component of calculating the KL divergence. If we had a way to consistently generate our desired distribution, we would already be done. If we happened to know our desired distribution exactly, we could either load it in "manually" (put gates in ourselves), or train the generator by itself using the KL divergence as a loss function. There may still be other ways to "extract" a loading algorithm from an opaque oracle that are simpler than QGANs, but as of right now no such method is widely known. 

Perhaps one could could train an "unloading" algorithm that takes the distribution from the oracle, and tries to "undo" it by returning the zero state. As per the above, since the target state is known one could train this system alone by gradient descent using the KL divergence as a loss function. Then, perhaps there is some computationally efficient way to convert the gates of this "unloading" algorithm into the gates of an inverse algorithm, which would "load" our state. This is an intriguing idea that we did not consider at the time, and would be an interesting direction for a future project. 

\section{Running a QGAN}

Now that all of the relevant features of QGAN training have been explained, we are ready to train our Bell-state into a circuit. Our initial state can be seen in figure 7. To do our training, we will use a training ratio of 5, which means that in each epoch we will do 5 discriminator updates and 1 generator update. To be fully precise, the 5 discriminator updates happen first, followed by the single generator update, but this is generally something that doesn't matter very much. We will then train over 100 epochs, meaning that we will do a total of 500 discriminator updates and 100 generator updates, evenly distributed across our whole training session. 

Here is the final state of our generator after training:

\begin{center}
    \includegraphics[scale = 0.35]{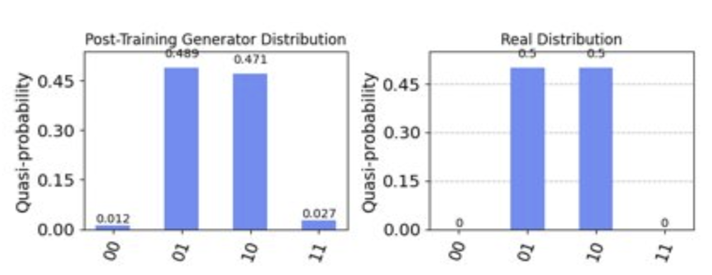}

    \emph{Fig 10: Final generator distribution (left) vs target distribution (right)}
\end{center}

Compared to our initial generator distribution (Fig 7), this is a significant improvement! Over the course of the training  run, it can be helpful for visualization if we plot the respective losses of our two circuits, and the KL divergence over the course of training:

\begin{center}
    \includegraphics[scale=0.7]{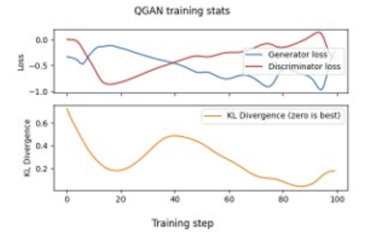}

    \emph{Fig 11: Training statistics for the 2-qubit Bell-state example}
\end{center}

The top graph shows the training loss of both the generator (blue) and discriminator (red) over the course of 100 epochs. The bottom graph shows the KL divergence over the course of 100 epochs. 

\section{Choosing Training Parameters}

This was a good proof of concept, but if we were doing this for real, how would we choose the training parameters, like the training ratio and number of epochs? If our choice of 5 and 100 respectively seem arbitrary, that's because they were. In general, the only way to pick the best values for training parameters is by brute trial and error. \emph{There is no known way to derive the optimal training parameters by hand from first-principles}. 

This is the inspiration for this research: if QGANs are to be used in the future for solving real problems, it would be far too inefficient to require quantum algorithm users to play around with training settings themselves. It would be much better if they could focus on the problem at hand, instead of worrying about choosing parameters. 

The objective is to test out many different parameter combinations, and pick out the best ones. The ones picked in the previous example according to our findings actually turn out to be pretty far from the best choice. Before talking about the main methods and findings of this research, we will demonstrate one more example problem, this time using 3 qubits and using the settings we found to be optimal. 

\section{3-Qubit Training Example}

Once again, we begin with a generator that is randomly initialized, and a target distribution, semi-randomly chosen by hand:

    \includegraphics[scale=0.4]{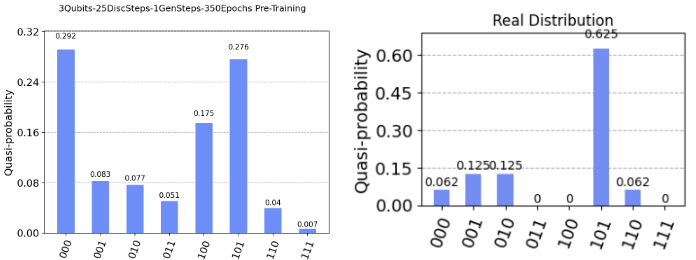}

    \emph{Fig 12: Initial and target distributions for the 3-qubit example}

We use 350 total epochs, with a training ratio of 25, as were found to be the best settings for 3-qubit distributions:

    \includegraphics[scale=0.6]{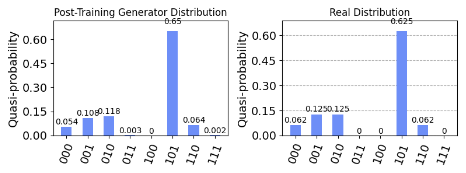}

    \includegraphics[scale=0.8]{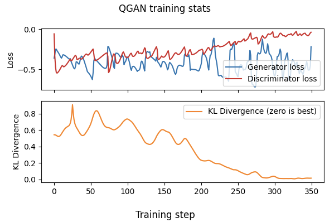}

    \emph{Fig 13: Final training statistics for our 3-qubit example}

3-qubit distributions are unsurprisingly much more expensive to learn than 2-qubit distributions. Computational complexity as a function of the number of qubits is a large focus of ours later on.
chapter{Testing and Results}

Now we will discuss the tests that were ran and the results that were found. As stated previously, our goal is to determine the optimal training parameters for learning an arbitrary n-qubit distribution. 

We set out to determine the best training ratio (discriminator updates to generator updates), and the best number of epochs as a function of the number of qubits. 

\section{Methods}

To do this, we will record both the average case, and best case performance (measured by KL-divergence, which we may not be able to see in real life, but it can still be used as a very helpful tool for us here, effectively allowing us to peek under the hood) across 25 completely randomized trials, using a wide variety of training ratios. 

In these trials, not only will the angles of our circuits be randomly initialized, but the target distribution will \emph{also} be randomly selected. It is important to ensure that the trials are completely randomized. Otherwise we could possibly run into some biases for the particular target-state that was chosen. 

For each qubit count, we generate 25 random trials. These same 25 trials are used for every training ratio that is tested. This is another strategy to remove possible biases in our data. If any distributions happen to be "easier" to learn than others, that effect should cancel out as we are only comparing relative performance.

We are unfortunately only able to run trials for 1, 2, and 3 qubit distributions. Any larger distributions would practically never run, since we are using a classical simulation of real quantum hardware. 

\section{Gathered Data}

Our gathered data is presented in the next six graphs, and will be explained shortly:

\begin{center}
    \includegraphics[scale=0.35]{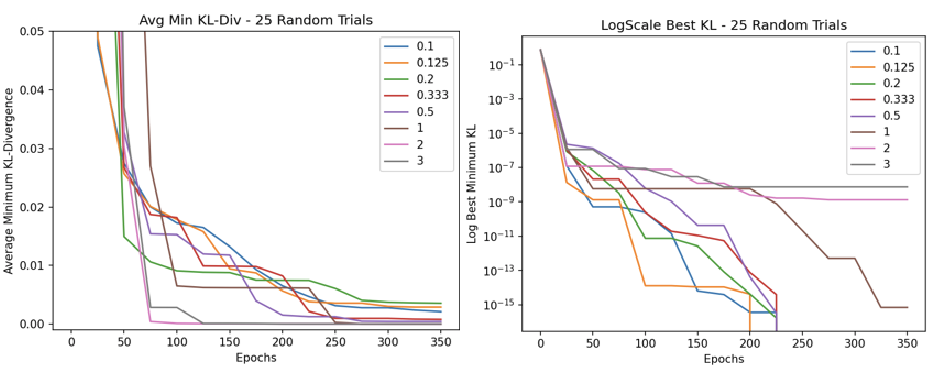}

    \includegraphics[scale=0.35]{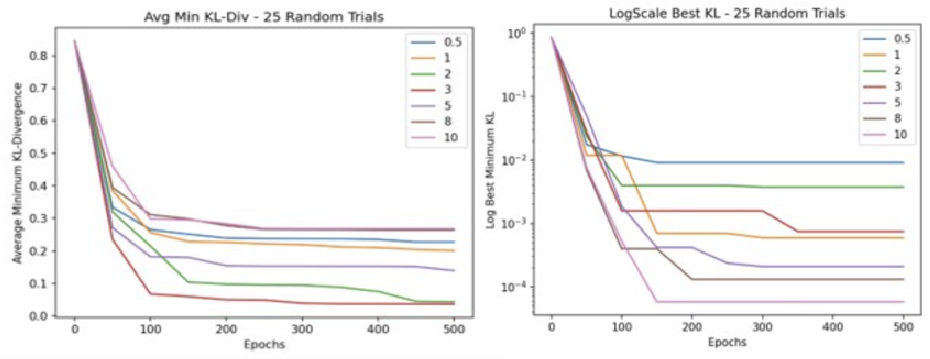}

    \includegraphics[scale=0.35]{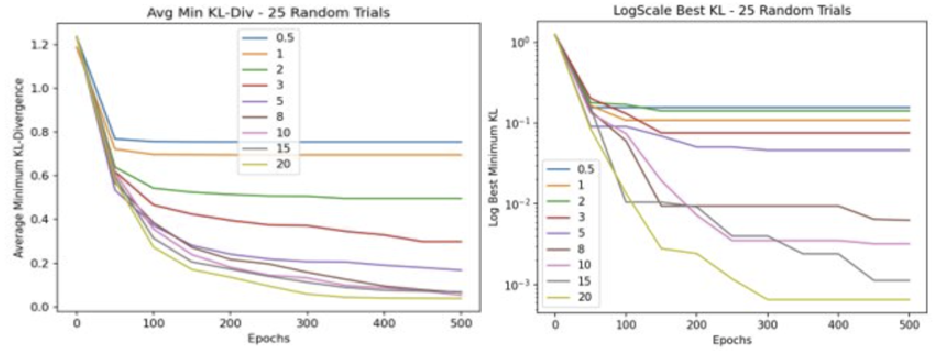}

    \emph{Fig 14: Data for 25 randomized trials for 1, 2, and 3 qubits.}
\end{center}

Our data for 1, 2, and 3 qubits appear in the top, middle, and bottom rows respectively. The left graphs show the average case performance, where as the right graphs show the best case performance. The x-axes are always the number of epochs, where as the y-axes are always the KL-Divergence. Different colored lines represent different training ratios, as indicated by the legends in each graph. 

As discussed previously, lower KL-divergence is better. Higher training ratios tend to achieve better performance almost across the board. The exception is for the best-case performance on 1-qubit distributions, where a training ratio of $\frac{1}{8}$ (1 discriminator update for every 8 generator updates) performs the best, and actually learns the distribution \emph{down to machine accuracy}! This can be seen in the top right graph by some of the lines spontaneously jumping down all the way to zero. 1-qubit distributions are the only setting where we are able to achieve this sort of precision, likely for reasons discussed above of our chosen circuit frameworks not being completely expressible. 

The following table is a summary of the best training settings found in the above graphs:

\begin{center}
    \includegraphics[scale=0.4]{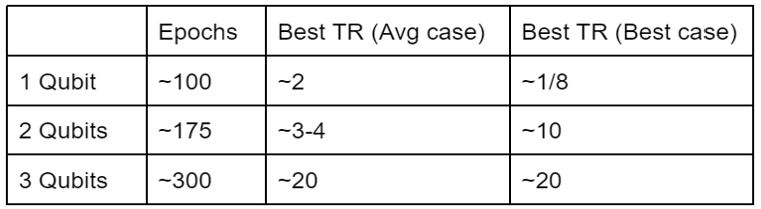}

    \emph{Fig 15: Table summary of the best training settings found for different qubit distributions.}
\end{center}

\subsection{Generalizing the Results}

We have done what we set out to do: if someone would like to use a QGAN to learn an arbitrary quantum distribution, they can use our circuit frameworks and pick the most efficient parameters from this table without having to do any testing themselves!

The problem is that if they desire to learn a distribution over more than 3 qubits, our data will be of no use to them, and they will be on their own in terms of picking efficient training parameters. Even if we had access to quantum hardware and were able to test much higher numbers of qubits, this would not change the fact that anyone who wants to go higher would be on their own.

Ideally, we'd like to generalize our results to numbers of qubits we have never tested before. We now attempt to do this using our admittedly limited data. 

The reason we are bothering at all is to try to answer a very important question: is the computational complexity of QGANs polynomial or exponential in terms of the number of qubits? Much of the inspiration for investigating QGANs in the first place was to combat the often exponential complexity of data-loading for quantum algorithms. If we were to find that our data suggests an exponential complexity, that could be terrible news for the future applicability of this technique. Thus we don't try to find exact relations to describe the optimal training parameters for an arbitrary number of qubits, but rather to find a good-enough relation to determine whether it seems to scale polynomially or exponentially.

\subsection{Testing with 4 Qubits}

An interesting way to do this is to use best-fit lines with our data so far, predict the best training parameters for 4 qubits, and run a small number of tests using 4-qubit distributions to check which best-fit line best predicts the new data point. First to find curves of best-fit for our gathered data:

\begin{center}
    \includegraphics[scale = 0.5]{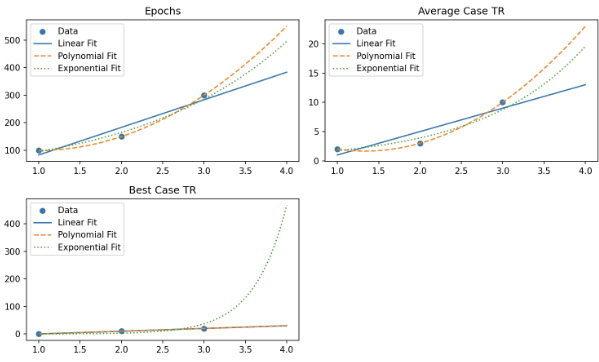}

    \emph{Fig 16: Linear (blue), Polynomial (orange), and Exponential (green) curves of best fit for our three categories of data: number of epochs, average case training ratio, and best case training ratio}
\end{center}

The best fit curves for the number of epochs and the average-case training ratio are fairly close together, and so they will only provide a small amount of information given the low sample size we will use. The best-case training ratio graph has a huge difference between the two best-fit curves however, and so the result here should be telling.

Now to run our 4-qubit test. There is a reason that we did not include 25 randomized 4-qubit trials with the rest of our data. One trial took my laptop over 50 hours for each set of parameters! We can only feasibly run one single trial, and 5 qubits would be entirely out of the question. 

Because we are only using 1 trial instead of 25, there is a chance our randomly-chosen distribution is not representative of an arbitrary distribution. Perhaps it could be easier, or more difficult, than an "average" distribution, and we have no way of knowing without running more trials, which is expensive. Nonetheless, here is the data we found for a random 4-qubit distribution, for a few different training ratios:

\begin{center}
    \includegraphics[scale=0.6]{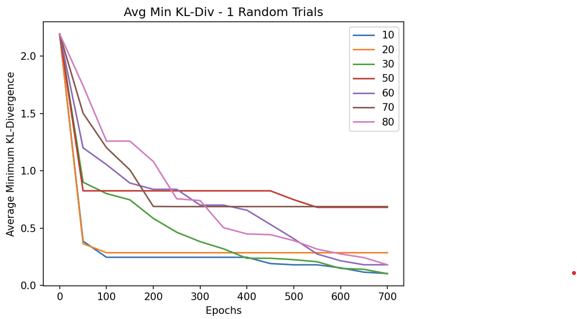}

    \emph{Fig 17: Training data for 4-qubits}
\end{center}

It appears that a training ratio of 30 does the best. Let's compare this to our curves of best fit:

\begin{center}
    \includegraphics[scale=0.7]{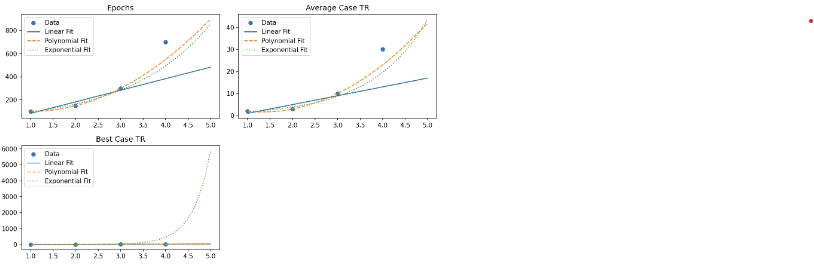}

    \emph{Fig 18: Curves of best fit with the new 4-qubit data point.}
\end{center}

In each case, our observed data point is better explained by the polynomial curve. This is great news! This means that QGANs are a feasible method for loading data in the future! We have found a solution to the initially posed problem: we have found a more efficient way to load data than standard exponential-complexity methods.

\section{Summary}

In this paper, we have built up some theoretical and mathematical tools for analyzing QGANs. We have investigated their performance in low-qubit settings. We have found some rudimentary estimates for optimal parameters for our chosen circuit frameworks, and have even shown that it is reasonable to expect these methods to scale well to higher-qubit settings. 

\section{Moving Forward}

There are lots of interesting directions to investigate beyond these results. Many have been discussed throughout the paper, but here we will discuss promising-looking research directions with the aim of helping future quantum algorithm users in practical applications.

Rather than predicting their needs and limiting factors, at this point in time it may be better to keep things as general as possible. The more assumptions about the future needs of quantum algorithm users that can be eliminated, the more likely it is that a given piece of research will remain helpful going forward.

The name of the game is costs and objectives. For instance, there are both classical and quantum computation costs. This requirement can be quantified by discussing the "relative costs" of quantum and classical computation. For instance, perhaps in terms of paying for electricity and renting out quantum hardware, "one" quantum computation is worth about 1000 classical computations. We'd then say that quantum computation is 1000x more expensive than classical computation. Given a relative cost rate between the two, there should be optimal settings which minimize the total overall cost. If such an analysis is done generally, with this cost rate factored out, your results would be useful in a much wider variety of possible futures as opposed to using the current relative cost rate. Research that does this would hold today, and would hold tomorrow even if a breakthrough in quantum computers were to completely invert the relative cost rate. 

Different users might also want different accuracy levels. Some users might want really precise solutions, where as others might be happier with less. In general, you can quantify this with a "utility function" over different accuracy levels. You could then find settings which maximize your total "gain" calculated as the difference between the utility of a given accuracy and the computational cost of achieving that accuracy. In other words: if you can spend a bit more computation to get a bit more accuracy, that extra compute better be worth less to you than that extra accuracy; spending more compute better be "worth" it to you, otherwise why bother?

Both of these facets can be quantified in terms of "prices". Research which finds the general relationship between optimal parameters and these prices would be very widely applicable, and would be incredibly helpful once quantum algorithms become more mainstream.

The other research direction that comes to mind relaxes yet another assumption: perhaps finding optimal parameters for learning arbitrary distributions is too general? In a practical application, it is not impossible that the distributions one might want to learn are not "arbitrary". Perhaps there are some distributions that are more likely than others, or even some that can be ruled out entirely. It is theoretically possible to use this information to reduce the computational costs of training, likely by choosing a suitable PQC framework. Doing so might be difficult, but findings in this direction could also be of large assistance to future quantum algorithm users.\\

\subsection{Conclusion}

Data loading is a worthwhile area of research for quantum algorithms. These problems will need to be solved one way or another in order to unlock the full range of benefit from quantum systems. QGANs seem to be a promising research direction, as is demonstrated by our results.

\bibliography{bibliography1}{}

\begin{thebibliography}{22}
\expandafter\ifx\csname natexlab\endcsname\relax\def\natexlab#1{#1}\fi
\expandafter\ifx\csname bibnamefont\endcsname\relax
  \def\bibnamefont#1{#1}\fi
\expandafter\ifx\csname bibfnamefont\endcsname\relax
  \def\bibfnamefont#1{#1}\fi
\expandafter\ifx\csname citenamefont\endcsname\relax
  \def\citenamefont#1{#1}\fi
\expandafter\ifx\csname url\endcsname\relax
  \def\url#1{\texttt{#1}}\fi
\expandafter\ifx\csname urlprefix\endcsname\relax\def\urlprefix{URL }\fi
\providecommand{\bibinfo}[2]{#2}
\providecommand{\eprint}[2][]{\url{#2}}

\bibitem[{\citenamefont{Dallaire-Demers and
  Killoran}(2018)}]{arxivQuantumGenerative}
\bibinfo{author}{\bibfnamefont{P.-L.} \bibnamefont{Dallaire-Demers}}
  \bibnamefont{and} \bibinfo{author}{\bibfnamefont{N.}~\bibnamefont{Killoran}},
  \emph{\bibinfo{title}{{Q}uantum generative adversarial networks ---
  arxiv.org}}, \bibinfo{howpublished}{\url{https://arxiv.org/abs/1804.08641}}
  (\bibinfo{year}{2018}).

\bibitem[{\citenamefont{Li et~al.}(2020)\citenamefont{Li, Zhang, and
  Xia}}]{LiZhang}
\bibinfo{author}{\bibfnamefont{T.}~\bibnamefont{Li}},
  \bibinfo{author}{\bibfnamefont{S.}~\bibnamefont{Zhang}}, \bibnamefont{and}
  \bibinfo{author}{\bibfnamefont{J.}~\bibnamefont{Xia}},
  \emph{\bibinfo{title}{Quantum generative adversarial network: A survey}},
  \bibinfo{howpublished}{Comput. Mater. Contin.64, 401–438}
  (\bibinfo{year}{2020}).

\bibitem[{\citenamefont{Biamonte and et. al.}(2017)}]{BiamonteWittek}
\bibinfo{author}{\bibfnamefont{J.}~\bibnamefont{Biamonte}} \bibnamefont{and}
  \bibinfo{author}{\bibfnamefont{P.~W.} \bibnamefont{et. al.}},
  \emph{\bibinfo{title}{Quantum machine learning}},
  \bibinfo{howpublished}{Nature, 549, 195–202} (\bibinfo{year}{2017}).

\bibitem[{\citenamefont{Ngo et~al.}(2023)\citenamefont{Ngo, Nguyen, and
  Thang}}]{Ngo}
\bibinfo{author}{\bibfnamefont{T.~A.} \bibnamefont{Ngo}},
  \bibinfo{author}{\bibfnamefont{T.}~\bibnamefont{Nguyen}}, \bibnamefont{and}
  \bibinfo{author}{\bibfnamefont{T.~C.} \bibnamefont{Thang}},
  \bibinfo{journal}{Electronics} \textbf{\bibinfo{volume}{12}}
  (\bibinfo{year}{2023}).

\bibitem[{\citenamefont{Luckin et~al.}(2016)\citenamefont{Luckin, Holmes,
  M.~Griffiths, and Forcer}}]{Holmes}
\bibinfo{author}{\bibfnamefont{R.}~\bibnamefont{Luckin}},
  \bibinfo{author}{\bibfnamefont{W.}~\bibnamefont{Holmes}},
  \bibinfo{author}{\bibfnamefont{M.}~\bibnamefont{M.~Griffiths}},
  \bibnamefont{and} \bibinfo{author}{\bibfnamefont{L.}~\bibnamefont{Forcer}},
  \emph{\bibinfo{title}{Intelligence Unleashed: An Argument for AI in
  Education}} (\bibinfo{publisher}{Pearson Education: London, UK},
  \bibinfo{year}{2016}).

\bibitem[{\citenamefont{Jeong and Yi}(2023)}]{Jeong}
\bibinfo{author}{\bibfnamefont{C.-H.} \bibnamefont{Jeong}} \bibnamefont{and}
  \bibinfo{author}{\bibfnamefont{M.~Y.} \bibnamefont{Yi}},
  \emph{\bibinfo{title}{Correcting rainfall forecasts of a numerical weather
  prediction model using generative adversarial networks}},
  \bibinfo{howpublished}{The Journal of Supercomputing volume 79, pages
  1289–1317} (\bibinfo{year}{2023}).

\bibitem[{\citenamefont{Besombes et~al.}(2021)\citenamefont{Besombes,
  Pannekoucke, Lapeyre, Sanderson, and Thual}}]{Besombes}
\bibinfo{author}{\bibfnamefont{C.}~\bibnamefont{Besombes}},
  \bibinfo{author}{\bibfnamefont{O.}~\bibnamefont{Pannekoucke}},
  \bibinfo{author}{\bibfnamefont{C.}~\bibnamefont{Lapeyre}},
  \bibinfo{author}{\bibfnamefont{B.}~\bibnamefont{Sanderson}},
  \bibnamefont{and} \bibinfo{author}{\bibfnamefont{O.}~\bibnamefont{Thual}},
  \bibinfo{journal}{Nonlinear Processes in Geophysics}
  \textbf{\bibinfo{volume}{28}}, \bibinfo{pages}{347} (\bibinfo{year}{2021}),
  \urlprefix\url{https://npg.copernicus.org/articles/28/347/2021/}.

\bibitem[{\citenamefont{Pantazi et~al.}(2016)\citenamefont{Pantazi, Moshou, and
  Bravo}}]{Pantazi}
\bibinfo{author}{\bibfnamefont{X.~E.} \bibnamefont{Pantazi}},
  \bibinfo{author}{\bibfnamefont{D.}~\bibnamefont{Moshou}}, \bibnamefont{and}
  \bibinfo{author}{\bibfnamefont{C.}~\bibnamefont{Bravo}},
  \emph{\bibinfo{title}{Active learning system for weed species recognition
  based on hyperspectral sensing}}, \bibinfo{howpublished}{Biosyst. Eng, 146,
  193–202} (\bibinfo{year}{2016}).

\bibitem[{\citenamefont{et. al}(2021{\natexlab{a}})}]{LussangeLazarevich}
\bibinfo{author}{\bibfnamefont{J.~L.} \bibnamefont{et. al}},
  \emph{\bibinfo{title}{Modelling stock markets by multi-agent reinforcement
  learning}}, \bibinfo{howpublished}{Comput. Econ., 57, 113–147}
  (\bibinfo{year}{2021}{\natexlab{a}}).

\bibitem[{\citenamefont{Hazra et~al.}(2016)\citenamefont{Hazra, Kumar, and
  Gupta}}]{HazraKumar}
\bibinfo{author}{\bibfnamefont{A.}~\bibnamefont{Hazra}},
  \bibinfo{author}{\bibfnamefont{S.}~\bibnamefont{Kumar}}, \bibnamefont{and}
  \bibinfo{author}{\bibfnamefont{A.}~\bibnamefont{Gupta}},
  \emph{\bibinfo{title}{Study and analysis of breast cancer cell detection
  using naïve bayes, svm and ensemble algorithms}},
  \bibinfo{howpublished}{Int. J. Comput. Appl. 145, 39-45}
  (\bibinfo{year}{2016}).

\bibitem[{\citenamefont{Djambic et~al.}(2016)\citenamefont{Djambic, Krajcar,
  and Bele.}}]{DjambicKracjar}
\bibinfo{author}{\bibfnamefont{G.}~\bibnamefont{Djambic}},
  \bibinfo{author}{\bibfnamefont{M.}~\bibnamefont{Krajcar}}, \bibnamefont{and}
  \bibinfo{author}{\bibfnamefont{D.}~\bibnamefont{Bele.}},
  \emph{\bibinfo{title}{Machine learning model for early detection of higher
  education students that need additional attention in introductory programming
  courses}}, \bibinfo{howpublished}{Int. J. Digit. Technol. Econ.1, 1–11}
  (\bibinfo{year}{2016}).

\bibitem[{\citenamefont{et. al}(2021{\natexlab{b}})}]{HuangDu}
\bibinfo{author}{\bibfnamefont{H.~H.} \bibnamefont{et. al}},
  \emph{\bibinfo{title}{Experimental quantum generative adversarial networks
  for image generation}}, \bibinfo{howpublished}{Phys. Rev. Appl. 16, 24051}
  (\bibinfo{year}{2021}{\natexlab{b}}).

\bibitem[{\citenamefont{Lloyd and Weedbrook}(2018)}]{arxivQuantumGenerative1}
\bibinfo{author}{\bibfnamefont{S.}~\bibnamefont{Lloyd}} \bibnamefont{and}
  \bibinfo{author}{\bibfnamefont{C.}~\bibnamefont{Weedbrook}},
  \emph{\bibinfo{title}{{Q}uantum generative adversarial learning ---
  arxiv.org}}, \bibinfo{howpublished}{\url{https://arxiv.org/abs/1804.09139}}
  (\bibinfo{year}{2018}).

\bibitem[{\citenamefont{Romero and
  Aspuru-Guzik}(2019)}]{arxivVariationalQuantum}
\bibinfo{author}{\bibfnamefont{J.}~\bibnamefont{Romero}} \bibnamefont{and}
  \bibinfo{author}{\bibfnamefont{A.}~\bibnamefont{Aspuru-Guzik}},
  \emph{\bibinfo{title}{{V}ariational quantum generators: {G}enerative
  adversarial quantum machine learning for continuous distributions ---
  arxiv.org}}, \bibinfo{howpublished}{\url{https://arxiv.org/abs/1901.00848}}
  (\bibinfo{year}{2019}).

\bibitem[{qis()}]{qiskitQuantumNeural}


\bibitem[{\citenamefont{Martyn et~al.}(2021)\citenamefont{Martyn, Rossi, Tan,
  and Chuang3}}]{arxivUnification}
\bibinfo{author}{\bibfnamefont{J.}~\bibnamefont{Martyn}},
  \bibinfo{author}{\bibfnamefont{Z.~M.} \bibnamefont{Rossi}},
  \bibinfo{author}{\bibfnamefont{A.~K.} \bibnamefont{Tan}}, \bibnamefont{and}
  \bibinfo{author}{\bibfnamefont{I.~L.} \bibnamefont{Chuang3}},
  \emph{\bibinfo{title}{A grand unification of quantum algorithms}},
  \bibinfo{howpublished}{\url{https://arxiv.org/abs/2105.02859}}
  (\bibinfo{year}{2021}).

\bibitem[{\citenamefont{Nielsen and Chuang}(2002)}]{Chuang}
\bibinfo{author}{\bibfnamefont{M.~A.} \bibnamefont{Nielsen}} \bibnamefont{and}
  \bibinfo{author}{\bibfnamefont{I.~L.} \bibnamefont{Chuang}},
  \emph{\bibinfo{title}{Quantum Computation and Quantum Information}}
  (\bibinfo{publisher}{MIT Press, Cambridge}, \bibinfo{year}{2002}).

\bibitem[{\citenamefont{Preskill}(2018)}]{Preskill}
\bibinfo{author}{\bibfnamefont{J.}~\bibnamefont{Preskill}},
  \emph{\bibinfo{title}{Quantum computing in the nisq era and beyond}},
  \bibinfo{howpublished}{Quantum, 2,79} (\bibinfo{year}{2018}).

\bibitem[{\citenamefont{Brassard et~al.}(2000)\citenamefont{Brassard, Hoyer,
  Mosca, and Tapp}}]{arxivQuantumAmplitude}
\bibinfo{author}{\bibfnamefont{G.}~\bibnamefont{Brassard}},
  \bibinfo{author}{\bibfnamefont{P.}~\bibnamefont{Hoyer}},
  \bibinfo{author}{\bibfnamefont{M.}~\bibnamefont{Mosca}}, \bibnamefont{and}
  \bibinfo{author}{\bibfnamefont{A.}~\bibnamefont{Tapp}},
  \emph{\bibinfo{title}{{Q}uantum {A}mplitude {A}mplification and {E}stimation
  --- arxiv.org}},
  \bibinfo{howpublished}{\url{https://arxiv.org/abs/quant-ph/0005055}}
  (\bibinfo{year}{2000}).

\bibitem[{\citenamefont{Harrow and Montanaro}(2017)}]{HarrowMontanaro}
\bibinfo{author}{\bibfnamefont{A.~W.} \bibnamefont{Harrow}} \bibnamefont{and}
  \bibinfo{author}{\bibfnamefont{A.}~\bibnamefont{Montanaro}},
  \emph{\bibinfo{title}{Quantum computational supremacy}},
  \bibinfo{howpublished}{Nature, 549, 203–209} (\bibinfo{year}{2017}).

\bibitem[{\citenamefont{Vint et~al.}(2021)\citenamefont{Vint, Anderson, Yang,
  Ilioudis, Di~Caterina, and Clemente}}]{VintAnderson}
\bibinfo{author}{\bibfnamefont{D.}~\bibnamefont{Vint}},
  \bibinfo{author}{\bibfnamefont{M.}~\bibnamefont{Anderson}},
  \bibinfo{author}{\bibfnamefont{Y.}~\bibnamefont{Yang}},
  \bibinfo{author}{\bibfnamefont{C.}~\bibnamefont{Ilioudis}},
  \bibinfo{author}{\bibfnamefont{G.}~\bibnamefont{Di~Caterina}},
  \bibnamefont{and} \bibinfo{author}{\bibfnamefont{C.}~\bibnamefont{Clemente}},
  \bibinfo{howpublished}{Remote Sensing 13, 596} (\bibinfo{year}{2021}).

\bibitem[{\citenamefont{Plesch and Časlav
  Brukner}(2011)}]{apsQuantumstatePreparation}
\bibinfo{author}{\bibfnamefont{M.}~\bibnamefont{Plesch}} \bibnamefont{and}
  \bibinfo{author}{\bibnamefont{Časlav Brukner}},
  \emph{\bibinfo{title}{{Q}uantum-state preparation with universal gate
  decompositions --- journals.aps.org}},
  \bibinfo{howpublished}{\url{https://journals.aps.org/pra/abstract/10.1103/PhysRevA.83.032302}}
  (\bibinfo{year}{2011}).

\end{thebibliography}
\end{document}